\documentclass[twocolumn,showpacs,preprintnumbers,amsmath,amssymb]{revtex4}

\usepackage{graphicx}
\usepackage{dcolumn}
\usepackage{bm}

\begin{document}

\title{Chirality-driven mass enhancement in the kagome Hubbard model}

\author{Masafumi Udagawa}
\author{Yukitoshi Motome}%
\affiliation{%
Department of Applied Physics, 
University of Tokyo, 
Tokyo 113-8656, Japan}%

\date{\today}

\begin{abstract}
We investigate the quasiparticle-mass enhancement 
in the Hubbard model on the frustrated kagome lattice 
by using a cluster extension of the dynamical mean-field theory. 
By analyzing the cluster density matrix, 
we find a hierarchy of energy scale among charge, spin, and chirality degrees of freedom. 
Large amount of entropy associated with the chirality is released at a much lower temperature 
than other energy scales for spin and charge fluctuations, 
leading to a sharp peak in the specific heat and the single-particle spectrum. 
The results manifest a generic mechanism of mass enhancement 
driven by an emergent composite degree of freedom 
under geometrical frustration. 
\end{abstract}

\maketitle

Heavy-fermion behavior is one of the most intriguing phenomena in correlated electron systems. 
Canonical examples are found in rare-earth compounds, 
in which a large quasiparticle-mass enhancement is observed, 
e.g., in the specific-heat coefficient $\gamma$. 
The origin has been argued on the basis of the Kondo effect ---
the screening of local {\it f}-electron
moments by conduction electrons~\cite{Kondo1964}. 
In this case, the local {\it f}-electron moments serve as an ``entropy reservoir" 
for the heavy-mass behavior~\cite{Hewson}.

Recently, a class of transition metal compounds has drawn considerable attention 
due to their heavy-fermion behavior.
A large $\gamma$ of the order of 100mJ/molK$^2$ was observed in 
several compounds such as
LiV$_2$O$_4$~\cite{Kondo97}, Y(Sc)Mn$_2$~\cite{Wada87}, and $\beta$-Mn~\cite{Shinkoda79}. 
The conventional Kondo scenario does not 
apply straightforwardly to this heavy-fermion behavior, 
since these $3d$-electron compounds do not have an obvious entropy reservoir 
like the localized $f$ moments in the rare-earth compounds. 
There have been many proposals for the origin of this puzzling behavior~\cite{Anisimov99,Eyert1999,Kusunose00,Tsunetsugu02,YamashitaUeda03,HattoriTsunetsugu09}, 
but the issue remains controversial so far.

One of the proposals is the effect of strong electron correlation under 
geometrical frustration of the lattice structure~\cite{Eyert1999,Tsunetsugu02,YamashitaUeda03}.  
Divergence of the quasiparticle mass at the correlation-driven Mott transition 
was firstly pointed out by Brinkman and Rice~\cite{Brinkman1970}, 
and later, a more sophisticated picture was provided 
by the dynamical mean-field theory (DMFT)~\cite{Georges96}. 
It was shown that the electron correlation results in separation of energy scale 
between charge and spin (and orbital) degrees of freedom of electrons: 
Namely, it suppresses charge fluctuations at an energy scale of the Coulomb repulsion, 
and leaves large spin (and orbital) fluctuations at much lower temperatures $T$, 
which serve as an entropy reservoir~\cite{Georges93}. 
This mass divergence, however, persists 
only when spatial correlations are neglected. 
Indeed, when DMFT is extended to include spatial correlations, 
an antiferromagnetic spin correlation develops at low $T$,
which suppresses the local spin fluctuations and collapses the heavy-fermion state~\cite{Zhang07}. 
It is widely considered that this obstacle can be circumvented 
by the geometrical frustration: 
The frustration suppresses the development of the spatial correlation 
and rejuvenates the masked heavy-fermion behavior. 
Hence, in this scenario, the heavy-fermion behavior is an intrinsic property 
associated with the criticality of the Mott transition, 
and the geometrical frustration plays a secondary role of uncovering 
it by suppressing spatial correlations. 

In this Letter, contrary to the prevailing view, 
we reveal an intensive role of the geometrical frustration 
on the quasiparticle-mass enhancement. 
We show that the frustration not only suppresses the spatial correlations 
but also brings into being a composite degree of freedom, 
which plays a role of the entropy reservoir. 
It is known that a composite object, 
such as spin chirality~\cite{Miyashita84} and 
self-organized clusters~\cite{Radaelli2002,Lee2002,Horibe2006}, 
often emerges in insulating systems, 
but we here explore an importance of such composites in correlated metals. 
By studying the kagome-lattice Hubbard model by a cluster extension of DMFT, 
we demonstrate that the spin chirality degree of freedom emerges 
even in the metallic region. 
We find that there is a hierarchy of energy scale among the charge, spin, and chirality, and that
the chirality is dominant at the lowest $T$ with carrying a large amount of entropy. 
The release of this chirality-associated entropy leads to the mass enhancement, 
with yielding a sharp peak in the specific heat as well as in the single-particle spectrum.

\begin{figure}[b]
\begin{center}
\includegraphics[width=0.45\textwidth]{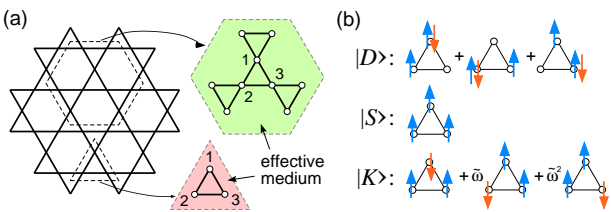}
\end{center}
\caption{\label{lattice_to_imp} 
(Color online) 
(a) Schematic picture of the mapping of the kagome lattice 
to three and nine-sites clusters in the CDMFT calculations. 
The three-site density matrix is calculated for the numbered three sites in each cluster. 
(b) Schematic pictures of the doublon, spin-polarized, and chiral states 
defined in Eqs.~(\ref{eq:D})-(\ref{eq:K}). 
}
\end{figure}

We consider the Hubbard model on the kagome lattice shown in Fig.~\ref{lattice_to_imp}, 
whose Hamiltonian is given by 
\begin{equation}
{\cal H} = -t \sum_{\langle ij \rangle,\sigma}
(c_{i\sigma}^\dagger c_{j\sigma} + {\text {h.c.}})
+ U \sum_i n_{i\uparrow} n_{i\downarrow},
\label{eq:H}
\end{equation}
where the sum of $\langle ij \rangle$ in the hopping term 
is taken over the nearest-neighbor sites, 
and $n_{i \sigma} = c_{i \sigma}^\dagger c_{i \sigma}$ 
in the onsite Coulomb repulsion term.  
The model has been extensively studied 
as a minimal model including both electron correlation and geometrical frustration
\cite{Imai03,Bulut05,Ohashi06,Bernhard07}.
Hereafter we set $t=1$ and the Boltzmann constant $k_{\rm B}=1$,
and fix the electron density at half filling.

We adopt the cellular dynamical mean-field theory (CDMFT)~\cite{Kotliar01} to study this model, 
which is an extension of DMFT to include the effect of spatial correlations within a cluster. 
The method is basically the same as that used in Ref.~\cite{Ohashi06}, however, 
in the present study, we perform the CDMFT calculations for two different clusters 
with three and nine sites (Fig.~\ref{lattice_to_imp}) 
to check the cluster-size dependence. 
Moreover, as a solver for the effective cluster model, 
we employ the continuous-time auxiliary-field quantum Monte Carlo (MC) method
based on the perturbation expansion in terms of $U$~\cite{Gull08}. 
This solver gives precise results 
more efficiently compared with the conventional Hirsch-Fye algorithm used in Ref.~\cite{Ohashi06}.
Typically we take 256 imaginary-time slices and $10^6$ MC steps, 
and the convergence is reached after 20 self-consistency loops, 
e.g., at $U=6$ and $T=0.1$. 
Larger computational cost is required at lower $T$ and larger $U$, 
in particular, in the calculations of the specific heat and the entropy (see below). 
The minus sign problem is not severe in the parameter range that we have investigated. 

In order to identify the relevant degrees of freedom in the system,
we calculate the cluster density matrix, 
which gives the probability distribution of quantum mechanical states within the cluster. 
Here we consider the density matrix defined on the three sites 
within each cluster 
indicated in Fig.~\ref{lattice_to_imp}. 
We calculate the diagonal components 
\begin{equation}
\rho_{\Psi} = \frac{1}{Z} \text{Tr} \left( |\Psi\rangle\langle\Psi| 
e^{- {\cal H}/T} \right)
\end{equation}
for a cluster state $|\Psi\rangle$, 
where $Z$ is the partition function. 
The $4^3=64$ states are classified by irreducible representations 
under the $U(1)\otimes SU(2)\otimes C_{3v}$ symmetry of the model. 
Among them, 
we focus on $\rho_{\Psi}$ for the following three states 
(and their symmetrically equivalent ones); 
\begin{eqnarray}
|D\rangle &=& \frac{1}{\sqrt{3}} \left( | \! 
\uparrow\downarrow, \uparrow, 0\rangle + | \, 0, \uparrow\downarrow, \uparrow\rangle 
+ | \! \uparrow, 0, \uparrow\downarrow\rangle \right), 
\label{eq:D}
\\
|S\rangle &=& | \! \uparrow,\uparrow,\uparrow\rangle, 
\label{eq:S}
\\
|K\rangle &=& \frac{1}{\sqrt{3}} \left( | \! 
\downarrow, \uparrow, \uparrow\rangle + \tilde{\omega} \, | \!
\uparrow, \downarrow, \uparrow\rangle + \tilde{\omega}^2| \! 
\uparrow, \uparrow, \downarrow\rangle \right),
\label{eq:K}
\end{eqnarray}
where 
$| \! \uparrow\downarrow, \uparrow, 0\rangle 
\equiv c_{1\uparrow}^{\dag}c_{1\downarrow}^{\dag}c_{2\uparrow}^{\dag}|\rm{vac}\rangle$, 
etc. ($|\rm{vac}\rangle$ is a vacuum), 
and $\tilde{\omega}$ is a phase factor related with the helicity (defined below). 
Here, we call $|D\rangle$ the doublon state which includes a doublon-holon pair,
$|S\rangle$ the spin-polarized state with the total spin $S=3/2$, and 
$|K\rangle$ the chiral state. 
Among these states, the chiral state is of special interest. 
$|K\rangle$ retains fourfold degeneracy with its time-reversal and reflection conjugate states: 
The four states are labeled by the two discrete quantum numbers, 
the $z$ component of total spin $S_z^{\rm tot} = \pm 1/2$ and the helicity $\eta = \pm 1$ 
[$\tilde{\omega}$ in Eq.~(\ref{eq:K}) is given by $\exp (i\frac{2}{3}\pi \eta)$]. 
We note that the chiral states are the eigenstates of 
scalar- and vector-chirality operators, studied 
in the previous CDMFT study~\cite{UdagawaICMproc}.

\begin{figure}[b]
\begin{center}
\includegraphics[width=0.4\textwidth]{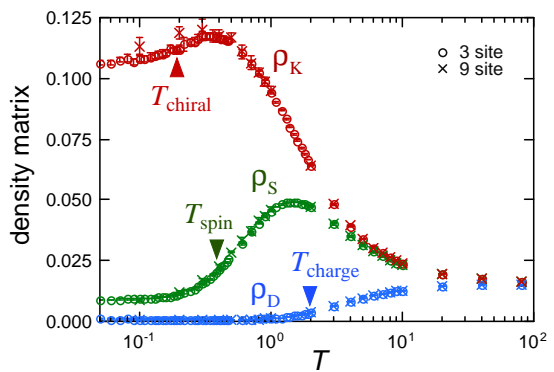}
\end{center}
\caption{\label{Global_dmatrix} 
(Color online)
Temperature dependences of the three-site density matrices, 
$\rho_D$, $\rho_S$, and $\rho_K$. 
The results for the three(nine)-site cluster are shown by circles (crosses).}
\end{figure}

Figure~\ref{Global_dmatrix} shows our CDMFT results for the density matrices, 
$\rho_D$, $\rho_S$, and $\rho_K$, 
in correlated metallic region at $U=6$. 
(The critical value for the Mott transition was estimated at $U_c \simeq 8.2$~\cite{Ohashi06}.) 
The three components show qualitatively different $T$ dependences.
At high $T$ enough, $\rho_D \approx \rho_S \approx \rho_K \approx 1/64$, 
since all the 64 states have nearly equal weights. 
As lowering $T$, $\rho_D$ is suppressed 
but $\rho_S$ and $\rho_K$ are enhanced. 
$\rho_S$ turns to decrease with showing a broad peak at $T \sim 1.2$. 
$\rho_K$ continues to increase down to lower $T$, 
but finally turns to decrease with showing a peak at $T \sim 0.32$. 
It is noteworthy that the results show little cluster-size dependence 
as shown in Fig.~\ref{Global_dmatrix}: 
This indicates that our CDMFT results 
show good convergence to the thermodynamic limit. 
Such rapid convergence may be attributed to 
suppressed inter-triangular correlations 
under the corner-sharing topology of the kagome lattice, 
as discussed in the localized spin models~\cite{Leung93,Moessner98}. 

These characteristic $T$ dependences indicate 
a hierarchy of energy scale for relevant degrees of freedom. 
Charge fluctuations are firstly suppressed by large $U$, 
being signaled by the suppression of $\rho_D$ which includes a doublon-holon pair. 
At a lower energy scale 
where $\rho_S$ decreases steeply, the spin degree of freedom is frozen out, 
and finally, the chirality is quenched at the lowest energy scale, 
corresponding to the suppression of $\rho_K$. 
We identify the characteristic energy scales 
for charge, spin, and chirality degrees of freedom by the temperatures 
$T_{\rm charge}$, $T_{\rm spin}$, and $T_{\rm chiral}$ where 
$\rho_D$, $\rho_S$, and $\rho_K$ are suppressed most rapidly 
(the inflection points indicated by the triangles in Fig.~\ref{Global_dmatrix}).
The hierarchy is clearly seen 
as $T_{\rm{charge}} \simeq 2.0 > T_{\rm{spin}} \simeq 0.37 > T_{\rm{chiral}} \simeq 0.18$. 

\begin{figure}[t]
\begin{center}
\includegraphics[width=0.45\textwidth]{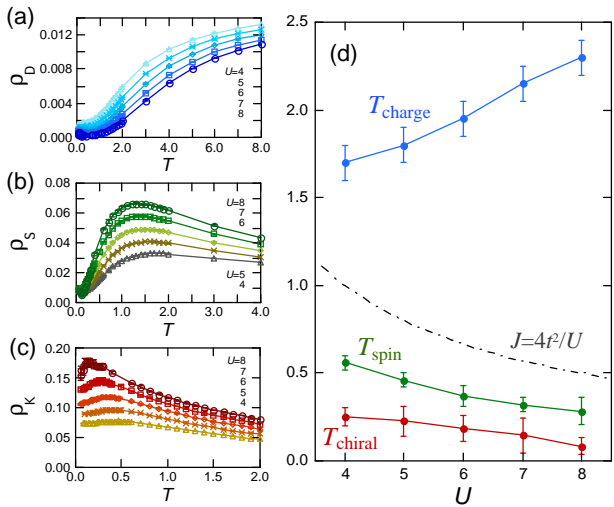}
\end{center}
\caption{\label{Crossover} 
(Color online) 
Temperature dependences of the density matrices (a) $\rho_D$, (b) $\rho_S$, and (c) $\rho_K$. 
The results are for the three-site cluster and 
the lines are guides for the eye.
The characteristic energy scales $T_{\rm{charge}}$,
$T_{\rm{spin}}$, and $T_{\rm{chiral}}$, determined 
by the steep decrease of $\rho_D$, $\rho_S$, and $\rho_K$, 
are plotted as a function of $U$ in (d). 
The dashed line shows the effective exchange interaction $J=4t^2/U$ for comparison. 
}
\end{figure}

We investigate the energy hierarchy systematically in the correlated metallic region. 
Figures~\ref{Crossover}(a)-(c) show 
$T$ dependences of $\rho_D$, $\rho_S$, and $\rho_K$ with varying $U$. 
The estimated characteristic temperatures are summarized 
in the crossover phase diagram in Fig.~\ref{Crossover}(d). 
$T_{\text{charge}}$ increases as $U$, which is reasonable since charge fluctuations 
are suppressed below $T \simeq U-W^*$ ($W^*$ is a renormalized band width). 
On the other hand, $T_{\text{spin}}$ decreases as $U$ increases. 
We compare $T_{\text{spin}}$ with an effective superexchange interaction $J = 4t^2/U$, 
derived in the strong-coupling expansion in $t/U$, 
and find approximately $T_{\text{spin}} \approx J/2$. 
This suggests that the reduction of $\rho_S$ is a strong-correlation effect, 
originating from the effective spin exchange 
under suppressed charge fluctuations. 
$T_{\text{chiral}}$ also decreases as $U$ increases, and 
is always located well below $T_{\rm{spin}}$. 
This indicates that the composite chiral degree of freedom is dominated 
by a much smaller energy scale compared with $J$.

\begin{figure}[t]
\begin{center}
\includegraphics[width=0.45\textwidth]{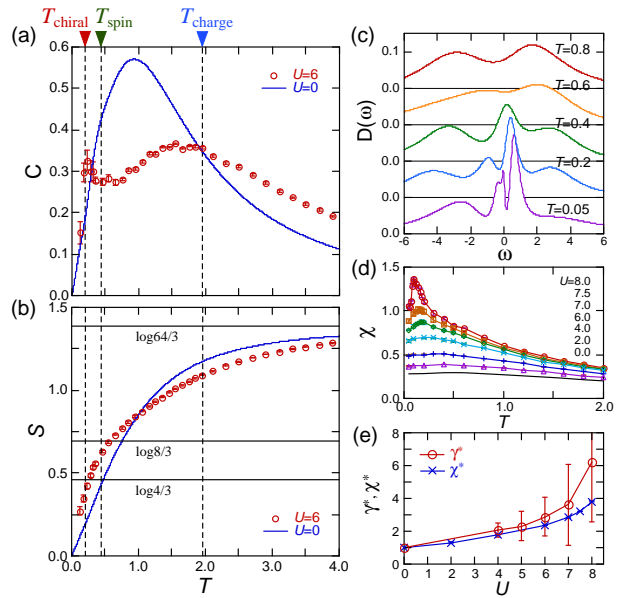}
\end{center}
\caption{\label{CandS} 
(Color online)
Temperature dependences of (a) specific heat and (b) entropy at $U=6$ per site. 
The results at $U=0$ are shown for comparison. 
Three vertical dashed lines indicate the characteristic temperatures estimated in Fig.~\ref{Crossover}, 
and three horizontal lines in (b) show the values of $\log4/3$, $\log8/3$, and $\log64/3$. 
(c) Single particle spectrum at $U=6$ obtained by the maximum entropy method. 
(d) Temperature dependence of the uniform magnetic susceptibility. 
(e) Enhancement factors of the specific-heat coefficient estimated from $T_{\text{chiral}}$, $\gamma^*$, 
and of the uniform magnetic susceptibility at $T=0.05$, $\chi^*$. See the text for details.
All the results are obtained by the three-site cluster DMFT. 
}
\end{figure}

The energy hierarchy and the quenching of each degree of freedom 
are also observed from the entropic point of view.
Figures~\ref{CandS}(a) and \ref{CandS}(b) show the CDMFT results for 
the specific heat $C$ and the entropy ${\cal S}$, respectively.  
Here, we calculated $C$ by numerical differentiation of the internal energy: 
To obtain the high precision data, 
we took a large number of imaginary-time slices, e.g., 
8192 slices at $T\simeq 0.1$ and $U\simeq 6$~\cite{note}. 
The entropy ${\cal S}$ is obtained by the numerical integration of $C/T$. 
As a result, $C$ shows a sharp peak at low $T \simeq T_{\text{chiral}}$ 
in addition to a broad peak at $T \sim T_{\text{charge}}$. 
The broad peak comes from the release of charge entropy; 
the peak temperature shifts to higher in accord with $T_{\text{charge}}$ as increasing $U$ (not shown).
On the other hand, the low-$T$ sharp peak originates from 
the entropy release associated with the chirality: 
In fact, as shown in Fig.~\ref{CandS}(b), 
the entropy $\cal{S}$ takes a value close to $\log 4 /3$ at $T_{\rm{chiral}}$, 
corresponding to the fourfold degeneracy of the chiral states $|K \rangle$. 
It is difficult to see a clear structure in $C$ at $T \simeq T_{\text{spin}}$, 
however, $\cal{S}$ reaches nearly $\log 8 /3$ at $T_{\text{spin}}$, 
reflecting the fourfold spin-polarized states $|S \rangle$ 
in addition to the four chiral states. 
Our results indicate that 
the entropy associated with chirality, spin, and charge degrees of freedom is released 
at separated energy scales 
$T_{\text{chiral}}$, $T_{\text{spin}}$, and $T_{\text{charge}}$ in Fig.~\ref{Crossover}(d).
To our knowledge, 
the resultant sharp peak in the specific heat due to the chirality 
has never been found in the previous studies.

The sharp peak is considered as a hallmark of the formation of Fermi liquid state 
below $T_{\rm{chiral}}$. 
In fact, as shown in Fig.~\ref{CandS}(c), 
the single-particle spectrum develops a sharp coherence peak at the Fermi level $\omega=0$ 
below $T_{\rm{chiral}}$.  
In addition, as plotted in Fig.~\ref{CandS}(d), 
the uniform magnetic susceptibility $\chi$ approaches a nonzero value as $T \to 0$ 
after showing a broad peak at around $T_{\rm{chiral}}$. 
Here $\chi$ is calculated from the magnetization 
by applying a small magnetic field. All the results consistently indicate that 
a Fermi liquid state emerges at low $T$ and 
the effective Fermi temperature is set by $T_{\rm{chiral}}$. 

In the low-$T$ Fermi liquid state, 
the quasiparticle mass is largely enhanced 
because the effective Fermi temperature $T_{\rm{chiral}}$ is suppressed. 
Although it is difficult to estimate the specific-heat coefficient $\gamma$ 
directly from the low-$T$ data of $C$, 
we obtain a rough estimate of $\gamma$ from the value of $T_{\text{chiral}}$:
Assuming that the chiral entropy $\log 4 /3$ is released 
uniformly in the range of $0 < T < T_{\rm{chiral}}$, 
we obtain the mass enhancement factor 
$\gamma^* = \gamma/\gamma_0 \simeq \log 4 / (3 T_{\rm{chiral}} \gamma_0) 
\simeq 3$ at $U=6$, 
where $\gamma_0 \simeq 0.873$ is the specific-heat coefficient at $U=0$.
As $T_{\rm{chiral}}$ is suppressed for larger $U$ [Fig.~\ref{Crossover}(d)], 
$\gamma^*$ is enhanced with increasing $U$, as shown in Fig.~\ref{CandS}(e). 
We also plot the enhancement factor $\chi^* = \chi/\chi_0$ 
estimated at $T=0.05$, where $\chi_0$ is the susceptibility at $U=0$.  
The results of $\gamma^*$ and $\chi^*$ 
consistently indicate a chirality-driven mass enhancement 
with approaching the Mott transition. 

In the formation of the heavy-quasiparticle state, 
the geometrical frustration plays an essential role, 
not only to suppress antiferromagnetic correlations 
but to bring about the degeneracy due to high local symmetry --- 
in the present model, 
the fourfold degeneracy of the chiral states in the triangular unit. 
In general, geometrically-frustrated lattices possess 
such local unit with high symmetry, e.g., triangles or tetrahedra, 
and hence, it is commonly expected that similar locally-degenerate state emerges. 
Our finding provides a prototypical example 
for the ubiquitous phenomenon in frustrated systems under strong correlation, 
that is, crossover from highly-symmetric local states to 
Fermi liquid state with an enhancement of quasiparticle mass.
We believe that the local degeneracy and emergent composite degree of freedom 
are of primary importance for the puzzling heavy-fermion state in transition metal compounds, 
such as LiV$_2$O$_4$~\cite{Kondo97}, Y(Sc)Mn$_2$~\cite{Wada87}, and $\beta$-Mn~\cite{Shinkoda79}. 

Finally let us remark on the fate of $T_{\text{chiral}}$ 
as the system enters Mott insulating region at larger $U$. 
In the sense that it sets the effective Fermi temperature, 
$T_{\text{chiral}}$ might terminate at the first-order Mott phase boundary. 
We note in fact that $T_{\text{chiral}}$ in Fig.~\ref{Crossover}(d) appear to approach 
the critical point ($U_c$, $T_c$) suggested in Ref.~\cite{Ohashi06}. 
On the other hand, the spin chirality itself 
can be important even in the insulating region: 
Higher order perturbations in terms of $t/U$ give higher-order exchange interactions 
in the effective spin model, which may induce
composite spin objects, such as the chirality. 
Unfortunately it is hard to clarify how $T_{\text{chiral}}$ behaves at $U>8$ 
because of the limitation of our simulation, 
and some complementary study is highly desired.
An interesting issue related to this problem is 
the recent experimental observation of a sharp peak in the low-$T$ specific heat of 
$^3$He on graphite at a commensurate filling $4/7$~\cite{Ishida97}. 
It has been argued that 
the system is insulating but in the vicinity of the Mott transition~\cite{Watanabe2007},
and the geometry of $^3$He atoms has 
frustration in between triangular and kagome lattices~\cite{Koretsune09}.

In summary, we have studied the correlated metallic region of 
the Hubbard model on the kagome lattice by the cluster dynamical mean-field theory, with the
continuous-time auxiliary-field quantum Monte Carlo method as an impurity solver.
We found that, 
in addition to charge and spin, 
the chirality degree of freedom becomes relevant in this system, and 
energy hierarchy appears among charge, spin, and chirality.
Moreover, we showed that the quasiparticle state with strongly renormalized mass is stabilized 
at low temperatures, as a manifestation of the strong frustration. 
The heavy-quasiparticle state is formed by the entropy release 
associated with the emergent composite degree of freedom, the chirality.
This chirality-driven mass enhancement provides a prototypical example
for the large mass behavior in frustrated systems under strong correlation, 
and gives a clue to controversies on the heavy-fermion behavior of transition metal compounds 
such as LiV$_2$O$_4$. 
It is interesting to extend the study for more realistic models 
by considering the effect of carrier doping and orbital degrees of freedom. 

The authors thank S.\ Sakai for fruitful discussions. We also thank S.\ Onoda and H.\ Tsunetsugu
for helpful comments. 
A part of the computations in this work has been done 
using the facilities of the Supercomputer Center, 
Institute for Solid State Physics, University of Tokyo.
This work was supported by Grants-in-Aid for Scientific Research on Priority Areas (Nos. 17071003 and 19052008), Grand-in-Aid for Young Scientists (B) (No. 21740242), Global COE Program ``the Physical Sciences Frontier", and the Next Generation Super Computing Project, Nanoscience Program, MEXT, Japan.

\end{document}